\documentstyle[11pt,aasms4]{article}
\singlespace

\def\kmsmpc{\,{\rm km\,s^{-1}\,Mpc^{-1}}}

\def\ldunits{\,{\rm h_{50}\,ergs\,s^{-1}\,Hz^{-1}\,Mpc^{-3}}}
\def\Haunits{\,{\rm ergs\,s^{-1}\,Mpc^{-3}}}
\def\lunits{\,{\rm ergs\,s^{-1}\,Hz^{-1}}}
\def\msun{\,{\rm M_\odot}}
\def\mden{\,{\rm h_{50}^2\,M_\odot\,Mpc^{-3}}}
\def\mdden{\,{\rm M_\odot\,Mpc^{-3}}}
\def\sfrd{\,{\rm M_\odot\,yr^{-1}\,Mpc^{-3}}}

\def\etal{{et al.\ }}
\def\ub{U-B}

\def\vi{V-I}

\def\spose#1{\hbox to 0pt{#1\hss}}
\def\lta{\mathrel{\spose{\lower 3pt\hbox{$\mathchar"218$}}
     \raise 2.0pt\hbox{$\mathchar"13C$}}}
\def\gta{\mathrel{\spose{\lower 3pt\hbox{$\mathchar"218$}}
     \raise 2.0pt\hbox{$\mathchar"13E$}}}
 
\begin{document}
\title{THE STAR FORMATION HISTORY OF FIELD GALAXIES}

\author{Piero Madau}

\affil{Space Telescope Science Institute, 3700 San Martin Drive,
Baltimore MD 21218;\ madau@stsci.edu}

\author{Lucia Pozzetti}

\affil{Dipartimento di Astronomia, Universit\`a di Bologna,
via Zamboni 33, I-40126 Bologna;\ lucia@astbo3.bo.astro.it}

\and

\author{Mark Dickinson\altaffilmark{1}$^,$\altaffilmark{2}}

\affil{Department of Physics and Astronomy, The Johns Hopkins University, 
Homewood Campus, Baltimore MD 21218;\ med@stsci.edu}

\altaffiltext{1}{Also at Space Telescope Science Institute}
\altaffiltext{2}{Allan C. Davis Fellow}

\begin{abstract}
We develop a method for interpreting faint galaxy data which focuses on the
integrated light radiated from the galaxy population as a whole. The emission
history of the universe at ultraviolet, optical, and near-infrared wavelengths
is modeled from the present epoch to $z\approx 4$ by tracing the evolution with
cosmic time of the galaxy luminosity density, as determined from several deep
spectroscopic samples and the {\it Hubble Deep Field} (HDF) imaging survey. In
a $q_0=0.5$, $h_{50}=1$ cosmology, the global spectrophotometric properties of
field galaxies can be well fit by a simple stellar evolution model, defined by
a time-dependent star formation rate (SFR) per unit comoving volume and a
universal initial mass function (IMF) extending from 0.1 to 125 $M_\odot$.
While a Salpeter IMF with a modest amount of dust reddening or a somewhat
steeper mass function, $\phi(m)\propto m^{-2.7}$, can both reproduce the data
reasonably well, a Scalo IMF produces too much long-wavelength light and is
unable to match the observed mean galaxy colors. In the best-fit models, the
global SFR rises sharply, by about an order of magnitude, from a redshift of
zero to a peak value  at $z\approx 1.5$ in the range 0.12--0.17 $\sfrd$, to
fall again at higher redshifts. After integrating the inferred star
formation rate over cosmic time, we find a stellar mass density at the present
epoch of $\Omega_sh_{50}^2\gta 0.005$, hence a mean stellar mass-to-light ratio
$\gta 4$ in the $B$-band and $\gta 1$ in $K$, consistent with the values
observed in nearby galaxies of various morphological types. The models are able
to account for the entire background light recorded in the galaxy counts down
to the very faint magnitude levels probed by the HDF. Since only $\sim 20$\% of
the current stellar content of galaxies is produced at $z>2$, a rather low
cosmic metallicity is expected at these early times, in good agreement with the
observed enrichment history of the damped Lyman-$\alpha$ systems. The biggest
uncertainty is represented by the poorly constrained amount of starlight that
was absorbed by dust and reradiated in the IR at early epochs. A ``monolithic
collapse'' model, where half of the present-day stars formed at $z>2.5$ and
were shrouded by dust, can be made consistent with the global history of light,
but overpredicts the metal mass density at high redshifts as sampled by QSO
absorbers. 

\end{abstract}


\section{Introduction}

In the past few years two different approaches have been widely used to
interpret faint galaxy data (see Ellis 1997 for a recent review). In the
simplest version of the ``traditional'' scheme, a one-to-one mapping between
galaxies at the present epoch and their distant counterparts is assumed: one
starts from the local measurements of the distribution of galaxies as a
function of luminosity and Hubble type and models their photometric evolution
assuming some redshift of formation and a set of parameterized star formation
histories (Tinsley 1980; Bruzual \& Kron 1980; Koo 1985; Guiderdoni \&
Rocca-Volmerange 1990; Metcalfe \etal 1991; Gronwall \& Koo 1995; Pozzetti,
Bruzual, \& Zamorani 1996). These, together with an initial mass function (IMF)
and a cosmological model, are then adjusted to match the observed number
counts, colors, and redshift distributions. Beyond the intrinsic simplicity of
assuming a well defined collapse epoch and pure-luminosity evolution
thereafter, the main advantage of this kind of approach is that it can easily
be made consistent with the {\it classical} view that ellipticals and spiral
galaxy bulges formed early in a single burst of duration 1 Gyr or less (see,
e.g. Ortolani \etal 1995 and references therein). Because much of the action
happens at high-$z$, however, these models predict far more Lyman-break ``blue
dropouts'' than are seen in the {\it Hubble Deep Field} (HDF) (Ferguson \&
Babul 1997; Pozzetti \etal 1997), and cannot reproduce the rapid evolution --
largely driven by late-type galaxies -- of the optical luminosity density with
lookback time observed by Lilly \etal (1996) and Ellis \etal (1996). Less
straighforward models which include, e.g., a large population of dwarf galaxies
that begin forming stars at $z\approx 1$ (Babul \& Ferguson 1996), or do not
conserve the number of galaxies due to merger events (Broadhurst, Ellis, \&
Glazebrook 1992; Carlberg \& Charlot 1993) also appear unable to match the
global properties of present-day galaxies (Ferguson 1997; Ferguson \& Babul
1997). 

A more physically motivated way to interpret the observations is to construct
semianalytic hierarchical models of galaxy formation and evolution (White \&
Frenk 1991; Lacey \& Silk 1991; Kauffmann \& White 1993; Kauffmann, White, 
\& Guiderdoni 1993; Cole \etal 1994; Baugh \etal 1997). Here,
one starts ab initio from a power spectrum of primordial density fluctuations,
follows the formation and merging of dark matter halos, and adopts various
prescriptions for gas cooling, star formation, feedback, and dynamical
friction. These are tuned to match the statistical properties of both nearby
and distant galaxies.  In this scenario, there is no period when bulges and
ellipticals form rapidly as single units and are very bright: rather, small
objects form first and merge continually to make larger ones. While reasonably
successful in recovering the counts, colors, and redshift distributions of
galaxies, a generic difficulty of such models is the inability to 
simultaneously reproduce the observed local luminosity density and the
zero-point of the Tully-Fisher relation (White \& Frenk 1991).

In this paper we shall develop an alternative method, which focuses on the emission
properties of the galaxy population {\it as a whole}. It traces the cosmic
evolution with redshift of the galaxy luminosity density -- as determined from
several deep spectroscopic samples and the HDF imaging survey -- and offers the
prospect of an empirical determination of the global star formation history of
the universe and initial mass function of stars independently, e.g., of the
merging histories and complex evolutionary phases of individual galaxies. The
technique relies on two basic properties of stellar populations: a) the
UV-continuum emission in all but the oldest galaxies is dominated by
short-lived massive stars, and is therefore a direct measure, for a given IMF
and dust content, of the instantaneous star formation rate (SFR); and b) the
rest-frame near-IR light is dominated by near-solar mass evolved stars that
make up the bulk of a galaxy's stellar mass, and can then be used as a tracer
of the total stellar mass density. By modeling the ``emission history'' of the
universe at ultraviolet, optical, and near-infrared wavelengths from the
present epoch to $z\approx 4$, we will shed light on some key questions in
galaxy formation and evolution studies: Is there a characteristic epoch of star
and metal formation in galaxies?  What fraction of the luminous baryons
observed today were already locked into galaxies at early epochs? Are high-$z$
galaxies obscured by dust? Do spheroids form early and rapidly? Is there a
universal IMF? 

Let us point out some of the limitations of our approach at the outset. (1)
We shall study the emission properties of ``normal'', optically-selected 
field galaxies which are only moderately affected by dust -- a typical spiral
emits 30\% of its energy in the far-infrared region (Saunders \etal 1990).
Starlight which is completely blocked from view even in the near-IR
by a large optical depth in dust will not be recorded by our technique, and the
associated baryonic mass and metals missed from our census. The contribution of
infrared-selected dusty starbursts to the present-day total stellar mass 
density cannot be very large, however, for otherwise the current limits to the
energy density of the mid- and far-infrared background would be violated (Puget
\etal 1996; Kashlinsky, Mather, \& Odenwald 1996; Fall, Charlot, \& Pei 1996;
Guiderdoni \etal 1997). Locally, infrared luminous galaxies are known to
produce only a small fraction of the IR luminosity of the universe (Soifer \&
Neugebauer 1991). (2) Our method bypasses the ambiguities associated with the
study of morphologically-distinct samples whose physical significance remains
unclear, but, by the same token, it does not provide any {\it direct}
information on the processes which shaped the Hubble sequence. Similarly, this
approach does not specifically address the evolution of particular subclasses
of objects, like the oldest ellipticals or low-surface brightness galaxies,
whose star formation history may have differed significantly from the global
average (e.g. Renzini 1995; McGaugh \& Bothun 1994). (3) Although in our
calculations the IMF extends from 0.1 to 125 $\msun$, by modeling the
rest-frame galaxy luminosity density from 0.15 to 2.2 \micron\ we will only be
sensitive to stars within the mass range from $\sim 0.8$ to about 20$\msun$.
This introduces non-negligible uncertainties in our estimate of the total
amount of stars and metals produced. (4) No attempt has been made to include
the effects of cosmic chemical evolution on the predicted galaxy colors. All
our population synthesis models assume solar metallicity, and thus will
generate colors that are slightly too red for objects with low metallicity,
e.g. truly primeval galaxies. (5) The uncertanties present in our estimates of
the UV luminosity density from the identification of Lyman-break galaxies in
the HDF are quite large, and the data points at $z>2$ should still be regarded
as tentative. This is especially true for the faint blue dropout sample at
$z\approx 4$, where  only one spectroscopic confirmation has been obtained so
far (Dickinson 1997). On the other hand, there is no evidence for a gross
mismatch at the $z\approx 2$ transition between the photometric redshift sample
of Connolly \etal (1997) and the Madau \etal (1996) UV dropout sample. 

The initial application of this method was presented by Lilly \etal (1996)
and Madau \etal (1996, hereafter M96). A complementary effort -- which starts
instead from the analysis of the evolving gas content and metallicity of the
universe -- can be found in Fall \etal (1996). Unless otherwise stated, we
shall adopt in the following a flat cosmology with $q_0=0.5$ and
$H_0=50\,h_{50}\kmsmpc$. 

\section{The Evolution of the Galaxy Luminosity Density}

The integrated light radiated per unit volume from the entire galaxy population
is an average over cosmic time of the stochastic, possibly short-lived star
formation episodes of individual galaxies, and will follow a relatively simple
dependence on redshift. In the UV -- where it is proportional to the global SFR
-- its evolution should provide information, e.g., on the mechanisms which
may prevent the gas within virialized dark matter halos from radiatively cooling
and turning into stars at early times, or on the epoch when galaxies exhausted
their reservoirs of cold gas. From a comparison between different wavebands it
should be possible to set constraints on the average IMF and dust content of
galaxies. 

The comoving luminosity density, $\rho_\nu(z)$, from the present epoch to
$z\approx 4$ is given in Table 1 in five broad passbands centered around 0.15,
0.28, 0.44, 1.0, and 2.2 \micron.  The data are taken from the $K$-selected
wide-field redshift survey of Gardner \etal (1997), the
$I$-selected  CFRS (Lilly \etal 1996) and $B$-selected Autofib (Ellis \etal
1996) surveys, the photometric redshift catalog for the HDF of Connolly \etal
(1997) -- which take advantage of deep infrared observations by Dickinson \etal
(1997) -- and the color-selected UV and blue ``dropouts'' of M96. \footnote{For
the UV dropouts, refined photometric criteria have been used after the many
redshift measurements with the Keck telescope (see Madau 1997a  and references
therein).}~ They have all been corrected for incompleteness by integrating over
the best-fit Schechter function in each redshift bin, 
\begin{equation}
\rho_\nu(z)=\int_0^\infty L_\nu\phi(L_\nu,z)dL_\nu=\Gamma(2+\alpha)\phi_*L_*.
\label{eq:ld} 
\end{equation}
As it is not possible from the Connolly \etal (1997) and M96 data sets 
to reliably determine the faint end slope of the luminosity function, a value of
$\alpha=-1.3$ has been assumed at each redshift interval for comparison with
the CFRS sample (Lilly \etal 1995). The error bars are typically less than 0.2
in the log, and reflect the systematic uncertainties introduced by the 
assumption of a particular value of $\alpha$ and, in
the HDF $z>2$ sample, in the volume normalization and color-selection region.
In the $K$-band, the determination by Gardner \etal (1997) agrees to within
30\% with Cowie \etal (1996), and we have assigned an error of 0.1 in the log
to the estimate of the local luminosity density at 2.2 \micron. 

Despite the obvious caveats due to the likely incompleteness in the data sets,
different selection criteria, and existence of systematic uncertainties in the
photometric redshift technique, the spectroscopic, photometric, and Lyman-break
galaxy samples appear to provide a remarkably consistent picture of the
emission history of field galaxies. The UV luminosity density rises sharply, by
about an order of magnitude, from a redshift of zero to a peak at $z\approx
1.5$, to fall again at higher redshifts (M96; Lilly \etal 1996; Connolly \etal
1997). This points to a rapid drop in the volume-averaged SFR in the last 8--10
Gyr, and to a redshift range $1\lta z\lta 2$ in which the bulk of the stellar
population was assembled. The decline in brightness at late epochs is shallower
at longer wavelengths, as galaxies becomes redder with cosmic time, on the
average. 

\section{Indicators of Past and Present Star Formation Activity}

Stellar population synthesis has become a standard technique to study the
spectrophotometric properties of galaxies. Here, we shall make extensive use of
the latest version of Bruzual \& Charlot (1993) isochrone synthesis code,
optimized with an updated library of stellar spectra (Bruzual \& Charlot 1997),
to predict the time change of the spectral energy distribution of a stellar
population. The uncertanties linked to the underlying stellar evolution
prescriptions and the lack of accurate flux libraries do not typically exceed
35\% (Charlot, Worthey, \& Bressan 1966). Shortward of the Lyman edge, however, 
the differences in the predicted ionizing radiation from model atmospheres of
hot stars can be quite large (Charlot 1996a). We shall consider three
possibilities for the IMF, $\phi(m)\propto m^{-1-x}$: a Salpeter (1955)
function ($x=1.35$), a Scalo (1986) function,  which is flatter for low-mass
stars and significantly less rich in massive stars than Salpeter, and an
intermediate case with $x=1.7$. In all models the metallicity is fixed to solar
values and the IMF is truncated at 0.1 and 125 $\msun$. 

\begin{figure}
\plotone{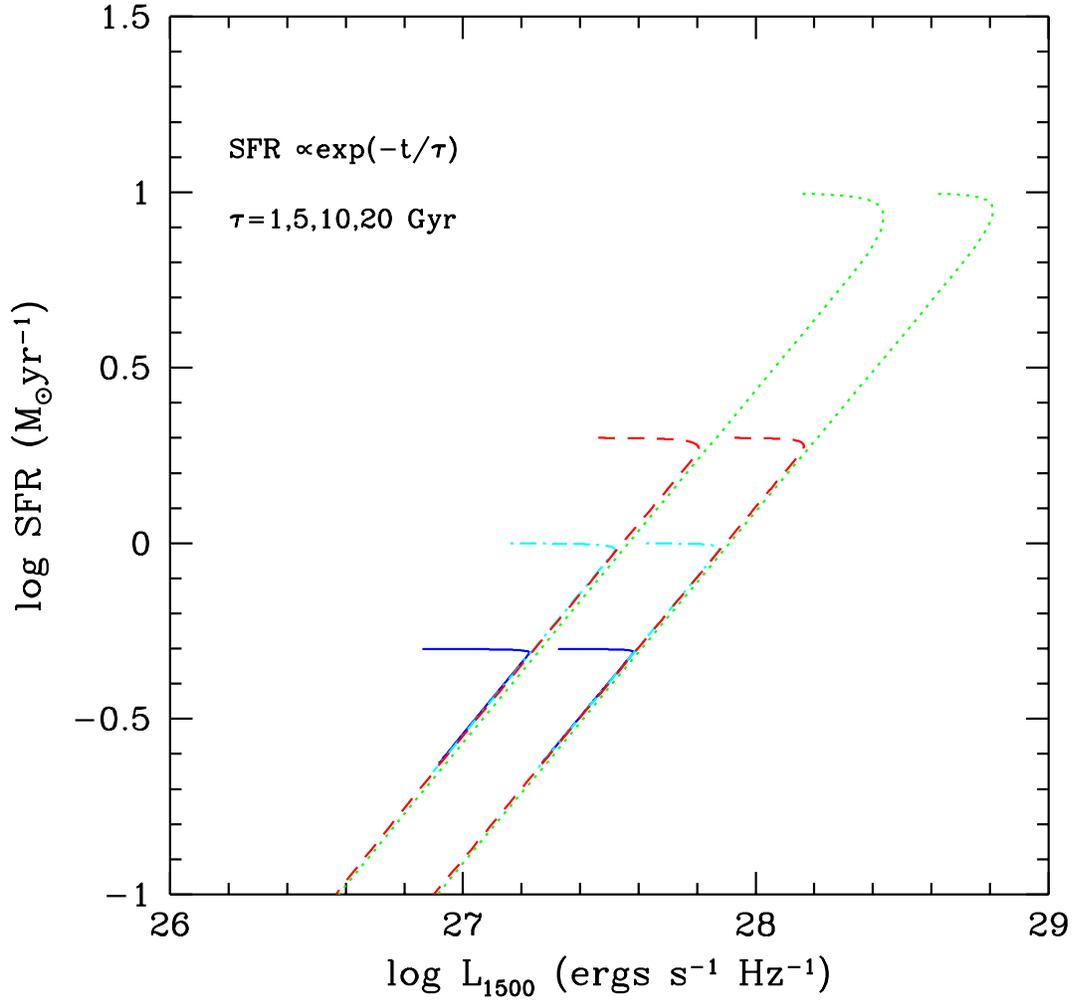}
\caption{SFR-UV relation for models with various exponentially
declining star formation rates at ages between 0.01 and 15 Gyr. {\it Solid
lines:} $\tau=20$ Gyr. {\it Dot-dashed lines:} $\tau=10$ Gyr. {\it Dashed
lines:} $\tau=5$ Gyr. {\it Dotted lines:} $\tau=1$ Gyr. The set of curves on
the left-hand side of the plot assume a Scalo IMF, the ones on the right-hand
side a Salpeter function.
\label{fig1}}
\end{figure}

\subsection{Birthrate--Ultraviolet Relation}

The UV continuum emission from a galaxy with significant ongoing star formation
is entirely dominated by late-O/early-B stars on the main sequence. As these
have masses $\gta 10\msun$ and lifetimes $t_{MS}\lta 2\times 10^7\,$yr, the
measured luminosity becomes proportional to the stellar birthrate and
independent of the galaxy history for $t\gg t_{MS}$. This is depicted in Figure
1, where the power radiated at 1500 \AA\ and 2800 \AA\ is plotted against the
instantaneous SFR for a model stellar population with different star formation
laws, SFR$\propto \exp(-t/\tau)$, where $\tau$ is the duration of the burst.
After an initial transient phase where the UV flux rises rapidly and the
turnoff mass drops below 10$\msun$, a steady state is reached where one can
write 
\begin{equation}
L_{UV}={\rm const}\times {{\rm SFR}\over \msun\, {\rm yr}^{-1}}\, \lunits,
\end{equation}
with const$=(8.0\times 10^{27}, 7.9\times 10^{27})$ at (1500\AA, 2800\AA) for a
Salpeter IMF, and const$=(3.5\times 10^{27}, 5.1\times 10^{27})$ for a Scalo
IMF, quite insensitive to the details of the past star formation
history.\footnote{The luminosities at 1500 \AA\ and 2800 \AA\ have been
averaged over a rectangular bandpass of width $\Delta\lambda/\lambda=20\%$ in
order to approximate the standard broadband filters used in the observations.}~
Note how, for burst durations $\lta 1 $Gyr and a Scalo IMF, the luminosity at
2800 \AA\ becomes a poor SFR indicator after a few $e$-folding times, when the
contribution of intermediate-mass stars becomes significant. After averaging
over the whole galaxy population, however, we will find that the UV continuum
is always a good tracer of the instantaneous rate of conversion of cold gas
into stars. 

\subsection{Mass--Infrared Relation}

If we assume a time-independent IMF, we can then use the results of stellar
population synthesis modeling, together with the observed UV emissivity, to
infer the evolution of the star formation activity in the universe (M96). As
already mentioned, the biggest uncertainty in this procedure is due to dust
reddening, as newly formed stars which are completely hidden by dust would not
contribute to the UV luminosity. The effect is potentially more serious at
high-$z$, as for a fixed observer-frame bandpass, one is looking further into
the ultraviolet with increasing redshifts. On the other hand, it should be
possible to test the hypothesis that star formation regions remain, on average,
largely unobscured by dust throughout much of galaxy evolution by looking at
the near-infrared light density. This will be affected by dust only in the most
extreme, rare cases, as it takes an $E(B-V)\gta4$ mag to produce an optical
depth of order unity at 2.2 \micron. 

Although different types of stars -- such as supergiants, AGB, and red
giants -- dominate the $K$-band emission at different ages in an evolving
stellar population, the mass-to-infrared light ratio is relatively insensitive
to the star formation history (Charlot 1996b). Figure 2 shows $M/L_K$ (in solar
units) as a function of age for models with various exponentially declining
SFR compared to the values observed in nearby galaxies of early to late 
morphological types.\footnote{The observations refer to the mass within the 
galaxy Holmberg radius, where the contribution by the dark matter
halo is expected to be small (Caldwell \& Ostriker 1981).}~ As the stellar
population ages, the mass-to-infrared light ratio remains very close to unity,
independent of the galaxy color and Hubble type. We can use this interesting
property to estimate the baryonic mass in galaxies from the local $K-$band
luminosity density, $\log \rho_K(0)=27.05\pm0.1\,\ldunits$ (Gardner \etal
1997). The observed range $0.6\,h_{50}\lta M/L_K\lta 1.9\,h_{50}$ translates
into a visible (stars$+$gas) mass density at the present day of 
\begin{equation} 
2\times 10^8 \lta \rho_{s+g}(0) \lta 6\times 10^8 \mden
\end{equation}
($0.003\lta \Omega_{s+g}\lta 0.009$). We shall see in the next section
how the observed integrated UV emission, with or without the addition of some
modest amount of reddening, may account for the bulk of the baryons traced by
the $K$-band light, and how initial mass functions with relatively few
high-mass stars (such as the Scalo IMF), or models with a large amount of dust
extinction at all epochs will tend to overproduce the near-infrared emissivity.

\begin{figure}
\plotone{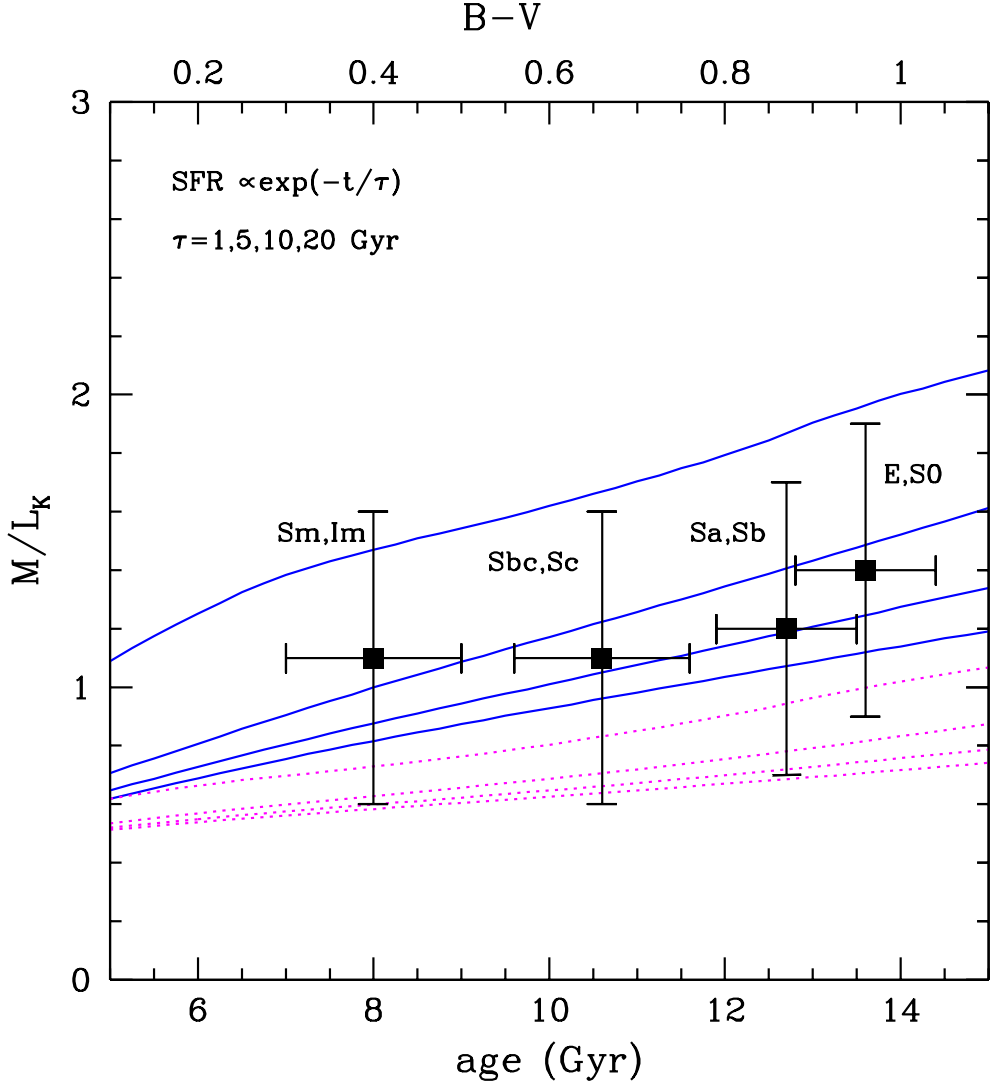}
\caption{Total (processed gas+stars) mass-to-$K$ band light ratio versus age
for models with various exponentially declining star formation rates. {\it
Solid lines:} Salpeter IMF. {\it Dotted lines:} Scalo IMF. From top to bottom,
each set of curves depict the values for $\tau=1, 5, 10,$ and 20 Gyr,
respectively. The data points show the visible mass-to-infrared light ratio
versus $B-V$ color (top axis) observed in nearby galaxies of various
morphological types (Charlot 1996b). 
\label{fig2}}
\end{figure}

\section{Stellar Population Synthesis Modeling}

An interesting question now arises as to whether a simple stellar evolution
model, defined by a time-dependent SFR per unit volume and a constant IMF, may
reproduce the global UV, optical, and near-IR photometric properties of the
universe as given in Table 1. In a stellar system with arbitrary star formation
rate, the luminosity density at time $t$ is given by the convolution integral 
\begin{equation}
\rho_\nu(t)=\int^t_0 L_\nu(\tau)\times {\rm SFR}(t-\tau)d\tau, 
\label{eq:rho} 
\end{equation}
where $L_\nu(\tau)$ is the specific luminosity radiated per unit initial mass
by a generation of stars with age $\tau$. In the instantaneous recycling
approximation (Tinsley 1980), the total stellar mass density produced at time
$t$ is 
\begin{equation}
\rho_s(t)=(1-R)\int_0^t {\rm SFR}(t)dt, 
\end{equation}
where $R$ is the mass fraction of a generation of stars that is returned to the
interstellar medium, $R\approx 0.3, 0.15,$ and 0.2 for a Salpeter, $x=1.7$, and
Scalo IMF, respectively. \footnote{To compute the return fraction $R$ we have
adopted the semiempirical initial-final mass relation of Weidemann (1987) for
stars with initial masses between 1 and 8 $\msun$. Stars with $M>10\msun$
have been assumed to return all but a $1.4\msun$ remnant.}~ In computing the
time evolution of the spectrophotometric properties of a stellar population in
comoving volumes large enough to be representative of the universe as a whole,
our first task is to relate the observed UV emission to a SFR density. We
assume a universal IMF and fit a smooth function to the UV continuum emissivity
at various redshifts. By construction, all models will then produce, to within
the errors, the right amount of ultraviolet light. We then use
Bruzual-Charlot's synthesis code to predict the cosmic emission history at long
wavelengths. 

In one of the scenarios discussed in the next section, the effect of dust
attenuation will be taken into account by multiplying equation (\ref{eq:rho})
by $p_{\rm esc}$, a time-independent term equal to the fraction of
emitted photons which are not absorbed by dust. For purposes of illustration,
we assume a foreground screen model, $p_{\rm esc}=\exp(-\tau_\nu)$, and
SMC-type dust. \footnote{Since what is relevant here is the absorption opacity,
we have multiplied the extinction optical depth by a factor of 0.6, as the
albedo of dust grains is known to approach asymptotically 0.4--0.5 at
ultraviolet wavelengths (e.g., Pei 1992).}~ This should only be regarded as an
approximation, since hot stars can be heavily embedded in dust within
star-forming regions, there will be variety of extinction laws, and 
the dust content of galaxies will evolve with redshift. While the existing data are too
sparse to warrant a more elaborate analysis, this simple approximation well
highlights the main features and assumptions of the model. It is possible to
gauge the luminosity-weighted amount of dust extinction at the present epoch by
looking at the observed local far-infrared luminosity density. For normal
galaxies, the emissivity from 8 to 115 \micron\ is estimated from the {\it
IRAS} survey to be around 30\% of their integrated emission in the $B$-band
(Saunders \etal 1990). Assuming that a negligible fraction of the ionizing flux
emerges in nebular lines, and a Salpeter IMF, this value implies a small
luminosity-weighted color excess, $E(B-V)\approx 0.012$ ($\tau_B\approx
0.025$). The Calzetti, Kinney, \& Storchi-Bergmann (1994) empirical extinction
law for starbursts, normalized to $A(V)/E(B-V)=4.88$, yields a {\it mean} dust
opacity that is similarly low. 

\subsection{Salpeter IMF}

\begin{figure}
\plotone{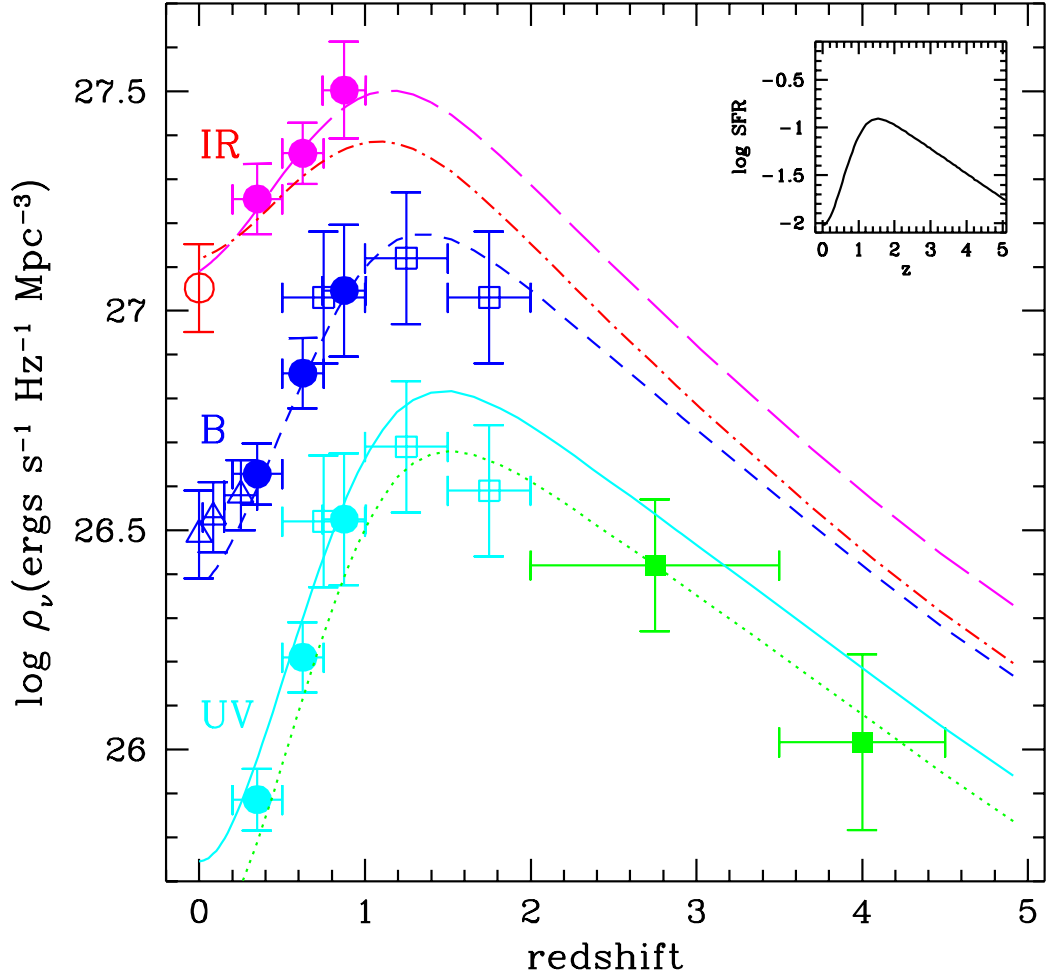}
\caption{Evolution of the luminosity density at rest-frame wavelengths of 0.15
({\it dotted line}), 0.28 ({\it solid line}), 0.44 ({\it short-dashed line}),
1.0 ({\it long-dashed line}), and 2.2 ({\it dot-dashed line}) \micron. The data
points with error bars are taken from Lilly \etal (1996) ({\it filled dots}
at 0.28, 0.44, and 1.0 \micron), Connolly \etal (1997) ({\it empty squares} at
0.28 and 0.44 \micron), Madau \etal (1996) and Madau (1997a) ({\it filled
squares} at 0.15 \micron), Ellis \etal (1996) ({\it empty triangles} at 0.44
\micron), and Gardner \etal (1997) ({\it empty dot} at 2.2 \micron). The inset
in the upper-right corner of the plot shows the SFR density ($\sfrd$) versus
redshift which was used as input to the population synthesis code. The model
assumes a Salpeter IMF, SMC-type dust in a foreground screen, and a universal
$E(B-V)=0.1$. 
\label{fig3}}
\end{figure}

Figure 3 shows the model predictions for the evolution of $\rho_\nu$ at
rest-frame ultraviolet to near-infrared frequencies. The data points with error
bars are taken from Table 1, and a Salpeter IMF has been assumed. In the
absence of dust reddening, this relatively flat IMF generates spectra that are
too blue to reproduce the observed mean (luminosity-weighted over the entire 
population) galaxy colors. The shape of
the predicted and observed $\rho_\nu(z)$ relations agrees better to within the
uncertainties if some amount of dust extinction, $E(B-V)=0.1$, is included.
In this case, the observed UV luminosities must be corrected upwards by a
factor of 1.4 at 2800 \AA\, and 2.1 at 1500 \AA. As expected, while the
ultraviolet emissivity traces remarkably well the rise, peak, and sharp drop in
the instantaneous star formation rate (the smooth function shown in the inset
on the upper-right corner of the figure), an increasingly large component of
the longer wavelengths light reflects the past star formation history. The peak
in the luminosity density at 1.0 and 2.2 \micron\ occurs then at later epochs,
while the decline from $z\approx 1$ to $z=0$ is more gentle than observed at
shorter wavelengths. The total stellar mass density at $z=0$ is
$\rho_s(0)=3.7\times 10^8\mdden$, with a fraction close to 65\% being produced
at $z>1$, and only 20\% at $z>2$.  In the assumed
cosmology, about half of the stars observed today are more than 9 Gyr old, and
only 20\% are younger than 5 Gyr.\footnote{Note that, unlike the measured
number densities of objects and rates of star formation, the integrated stellar
mass density does not depend on the assumed cosmological model.} 

\subsection{x=1.7 IMF}

Figure 4 shows the model predictions for a $x=1.7$ IMF and negligible dust
extinction. While able to reproduce quite well the $B$-band emission history
and consistent within the error with the local $K$-band light, this model
slightly underestimates the 1 \micron\ luminosity density at $z\approx 1$. The
total stellar mass density today is larger than in the previous case,
$\rho_s(0)=6.2\times 10^8\mdden$. 

\begin{figure}
\plotone{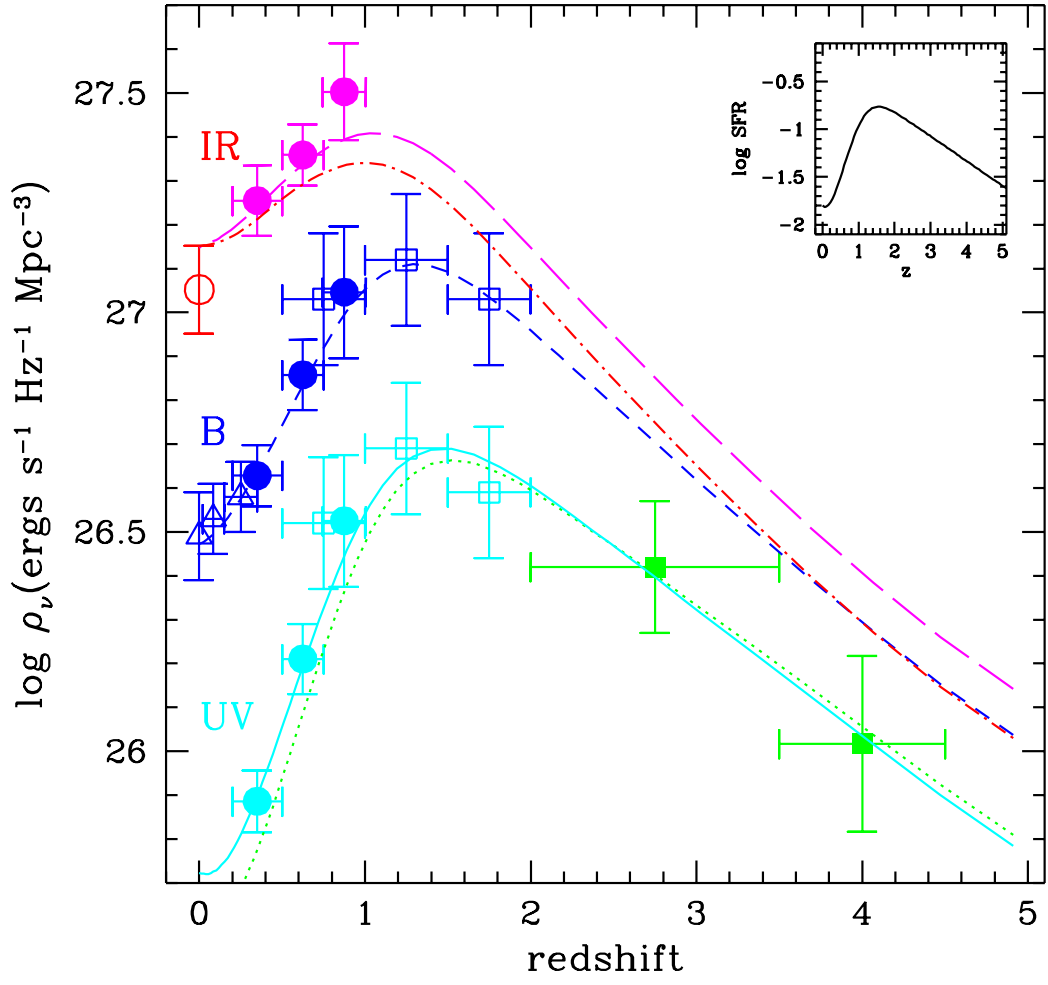}
\caption{Same as in Figure 3, but assuming an IMF with $\phi(m)\propto
m^{-1-1.7}$ and no dust extinction. 
\label{fig4}}
\end{figure}

\subsection{Scalo IMF}

Figure 5 shows the model predictions for a Scalo IMF. The fit to
the data is now much poorer, since this IMF generates spectra that are too red
to reproduce the observed mean galaxy colors, as first noted by Lilly \etal
(1996). Because of the relatively large number of solar mass stars formed, it
produces too much long-wavelength light by the present epoch. The addition of
dust reddening would obviously make the fit even worse. The total stellar mass
density produced is similar to the Salpeter IMF case. 

\begin{figure}
\plotone{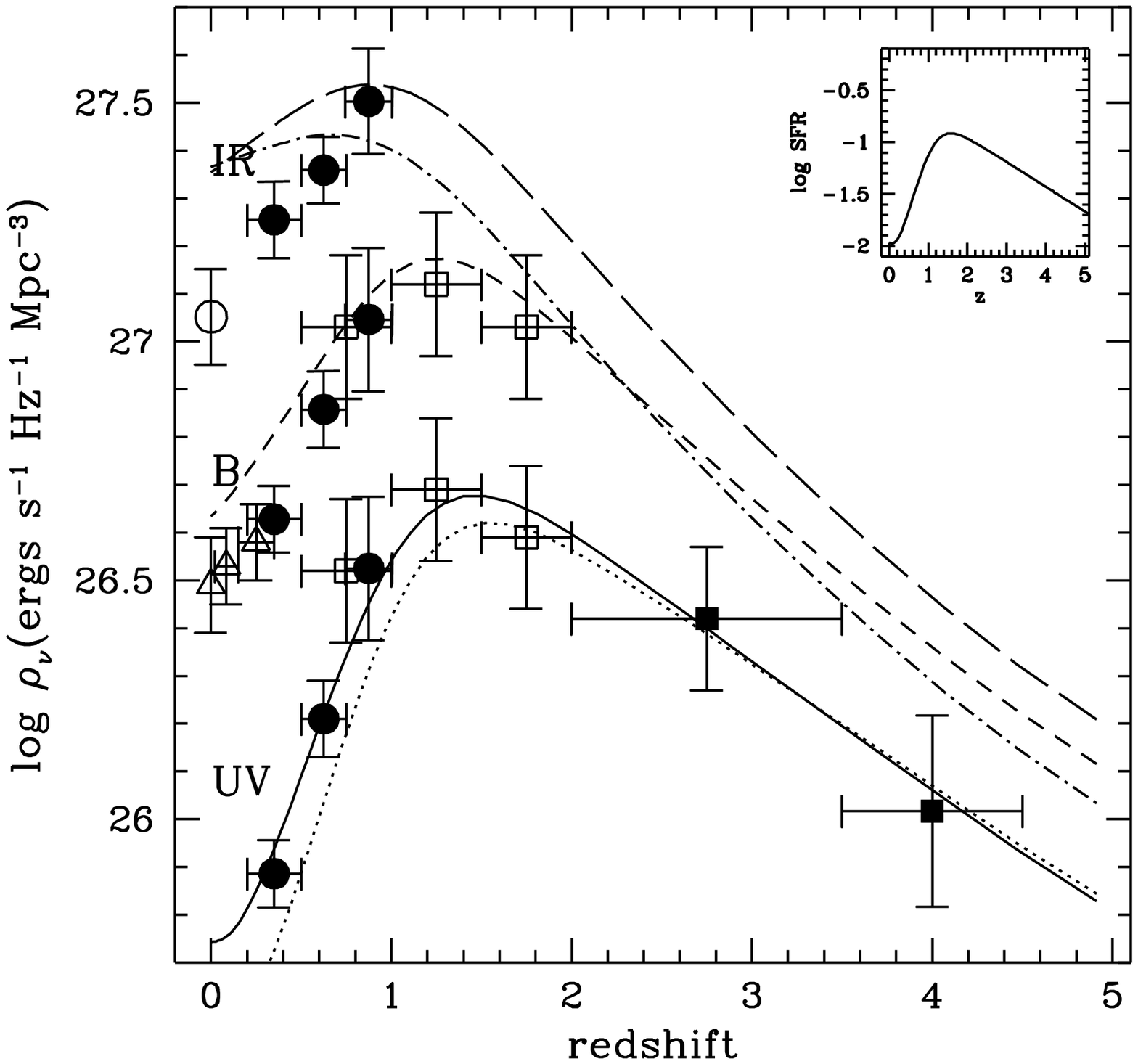}
\caption{Same as in Figure 3, but assuming a Scalo IMF and no dust extinction.
This model overproduces the local $K$-band emissivity by a factor of 2.
\label{fig5}}
\end{figure}

\section{Clues to Galaxy Formation and Evolution}

The results shown in the previous section have significant implications for our
understanding of the global history of star and structure formation. Here we
discuss a few key issues which will assist in interpreting the evolution of
luminous matter in the universe. 

\subsection{Extragalactic Background Light}

Our modeling of the data points to the redshift range where the bulk of the
stellar mass was actually produced: $1\lta z\lta 2$. The uncertainties in the
determination of the luminosity density at that epoch are, however, quite
large. At $z\approx 1$, the increase in the ``estimated'' emissivity 
(i.e., corrected for incompleteness by integrating over the best-fit Schechter
function) over that ``directly'' observed in the CFRS galaxy sample is about 
a factor of 2 (Lilly \etal 1996). Between $z=1$ and $z=2$, the peak in the
average SFR is only constrained by the photometric redshifts of Connolly \etal
(1997) and by the HDF UV dropout sample, both of which may be subject to
systematic biases. 

An important check on the inferred emission history of field galaxies
comes from a study of the extragalactic background light (EBL), an indicator of
the total optical luminosity of the universe. The contribution of known galaxies
to the EBL can be calculated directly by integrating the emitted flux times the
differential galaxy number counts down to the detection threshold. We have used
a compilation of ground-based and HDF data down to very faint magnitudes
(Pozzetti \etal 1997; Williams \etal 1996) to compute the mean surface
brightness of the night sky between 0.35 and 2.2 \micron. 
The results are plotted in Figure 6, along with the EBL spectrum
predicted by our modeling of the galaxy luminosity density, 
\begin{equation}
J_\nu={1\over 4 \pi}\int_0^\infty dz {dl\over dz}\rho_{\nu'}(z)
\end{equation}
where $\nu'=\nu(1+z)$ and $dl/dz$ is the cosmological line element. The overall
agreement is remarkably good, with the model spectra being only slightly bluer,
by about 20--30\%, than the observed EBL. The straightforward conclusion of
this exercise is that {\it the star formation histories depicted in Figures 3
and 4 appear able to account for the entire background light recorded in the
galaxy counts down to the very faint magnitudes probed by the HDF.} 

\begin{figure}
\plotone{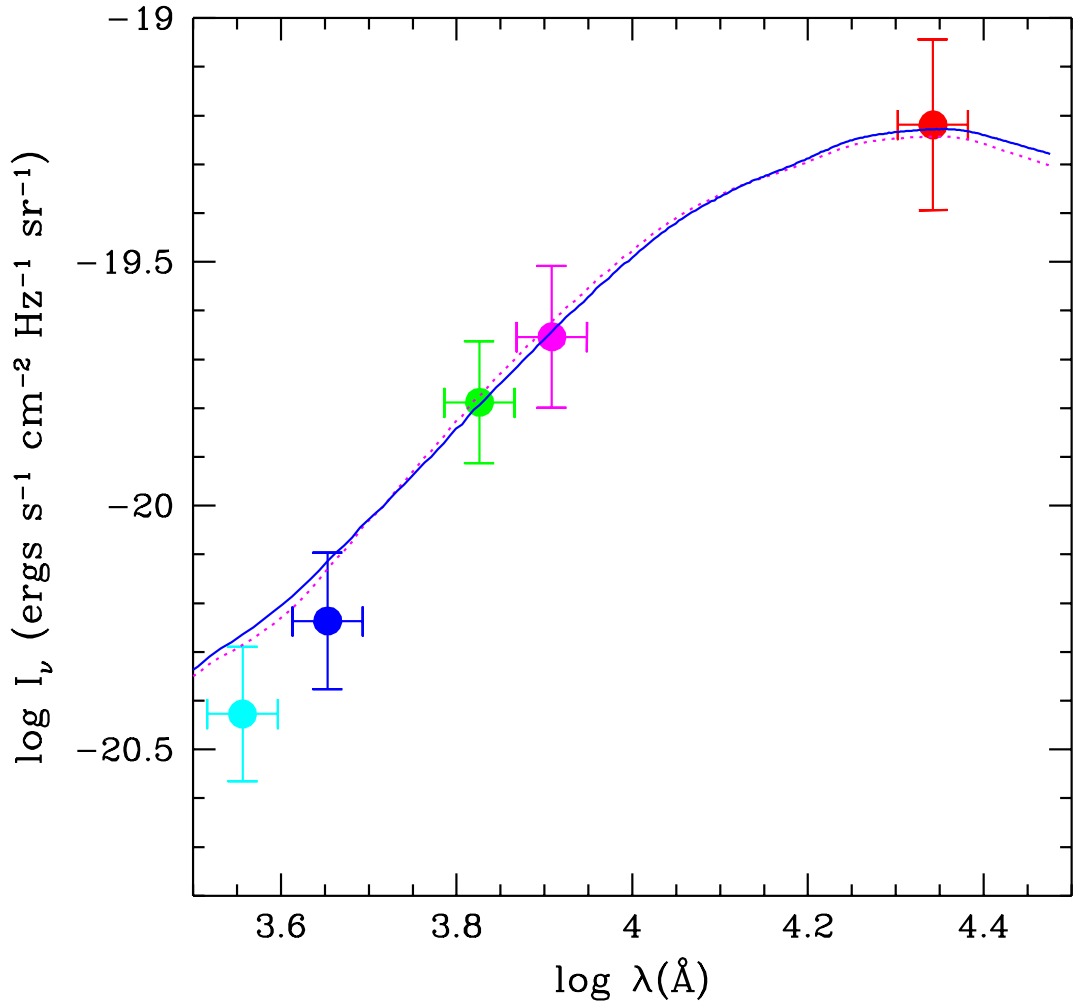}
\caption{Spectrum of the extragalactic background light as derived from a 
compilation of ground-based and HDF galaxy counts (see Pozzetti \etal 1997).
The 2$\sigma$ error bars arise mostly from field-to-field variations. {\it
Solid line:} Model predictions for a Salpeter IMF and $E(B-V)=0.1$ (star 
formation history of Figure 3). {\it Dotted line:} Model predictions for 
a $x=1.7$ IMF and negligible dust extinction (star formation history of Figure
4). 
\label{fig6}} 
\end{figure}

\subsection{The Stellar Mass Density Today}

The best-fit models discussed in \S\,4 generate a present-day stellar mass
density in the range between 4 and 6 $\times 10^8\mdden$
($0.005\lta\Omega_sh_{50}^2\lta 0.009$). Although one could in principle reduce
the inferred star formation density by adopting a top-heavy IMF, richer in
massive UV-producing stars, in practice a significant amount of dust reddening
-- hence of ``hidden'' star formation -- would then be required to match the
observed galaxy colors. The net effect of this operation would be a large
infrared background (see below). The stellar mass-to-light ratios range from
4.5 in the $B$-band and 0.9 in $K$ for a Salpeter function, to 8.1 in $B$ and
1.5 in $K$ for a $x=1.7$ IMF. Note that these values are quite sensitive to the
lower-mass cutoff of the IMF, as very-low mass stars can contribute
significantly to the mass but not to the integrated light of the whole stellar
population. A lower cutoff of 0.2$\msun$ instead of the 0.1$\msun$ adopted
would decrease the mass-to-light ratio by a factor of 1.3 for a Salpeter
function, 1.6 for $x=1.7$, and 1.1 for a Scalo IMF. 

\subsection{The Star Formation Density Today}

The predicted local rate of star formation ranges between 0.95$\times 10^{-2}$
(Salpeter) and 1.6$\times 10^{-2}\sfrd$ ($x=1.7$). According to Gallego \etal
(1995), the H$\alpha$ luminosity density of the local universe is $\log
\rho_{\rm H\alpha}=39.1\pm0.04\Haunits$. Let us assume case-B recombination
theory, and adopt the mean {\it unreddened} spectrum at $z=0$ corresponding to
the Salpeter IMF model, i.e., assume the emitted ionizing photons are 
converted to H$\alpha$ radiation before being absorbed by dust. The H$\alpha$
luminosity density can then be related to the local SFR per unit volume
according to 
\begin{equation}
\log \rho_{\rm H\alpha}=39.2 +\log\left({{\rm SFR}\over 0.01 \sfrd}\right)
\Haunits, 
\end{equation} 
(the coefficient for the $x=1.7$ case is 38.6) and is in good agreement with
the Gallego \etal determination. Very recently, we became aware of the first
results of an ongoing redshift survey of galaxies imaged in the rest-frame
ultraviolet at 2000 \AA\ with the FOCA balloon-borne camera (Treyer \etal 1997;
Milliard \etal 1992). The light density resulting from the integrated emission
of the observed galaxies, which have mean redshift $\langle z \rangle=0.15$, 
is $\rho_{2000}=8.0\pm 0.35\times 10^{25} \ldunits$, again in reasonable
agreement with the predicted value of $\approx 5\times 10^{25} \ldunits$. 

\subsection{Star Formation at High Redshift: Monolithic Collapse Versus 
Hierarchical Clustering Models} 

We have already cautioned the reader against the significant uncertanties present
in our estimates of the star formation density at $z>2$. The biggest  one is
probably associated with dust reddening, but, as the color-selected HDF sample
includes only the most actively star-forming young objects, one could also
imagine the existence of a large population of relatively old or faint galaxies
still undetected at high-$z$. The issue of the amount of star formation at
early epochs is a non trivial one, as the two competing models, ``monolithic
collapse'' versus hierarchical clustering, make very different predictions in
this regard. From stellar population studies we know in fact that about half of
the present-day stars are contained in spheroidal systems, i.e., elliptical
galaxies and spiral galaxy bulges (Schechter \& Dressler 1987). In the
monolithic scenario these formed early and rapidly, experiencing a bright
starburst phase at high-$z$ (Eggen, Lynden-Bell, \& Sandage 1962; Tinsley \&
Gunn 1976; Bower \etal 1992). In hierarchical clustering theory, instead, 
objects form and grow throughout the history of the universe by a process
of mergers and accretion (White \& Frenk 1991; Searle \& Zinn 1978). In such
models, elliptical galaxies form late by mergers of roughly equal mass disks
(Kauffmann et al. 1993), and most galaxies never experience star formation
rates in excess of a few solar masses per year (Baugh \etal 1997). The star
formation histories discussed in \S~4 produce only 20\% of the current stellar
content of galaxies at $z>2$, in apparent agreement with hierarchical
clustering theories. In fact, the tendency to form the bulk of the stars at
relatively low redshifts is a generic feature not only of the $\Omega_0=1$ CDM
cosmology, but also of successful low-density CDM models (cf Figure 21 of Cole
\etal 1994; Baugh \etal 1997). 

It is then of interest to ask how much larger could the volume-averaged SFR at
high-$z$ be before its fossil records -- in the form of long-lived, near
solar-mass stars -- became easily detectable as an excess of $K$-band light at
late epochs. In particular, is it possible to envisage a toy model where 50\%
of the present-day stars formed at $z>2.5$ and were shrouded by dust? The
predicted emission history from such a scenario is depicted in Figure 7. To
minimize the long-wavelength emissivity associated with the radiated
ultraviolet light, a Salpeter IMF has been adopted. Consistency with the HDF
data has been obtained assuming a dust extinction which increases rapidly with
redshift, $E(B-V)=0.011(1+z)^{2.2}$. This results in a correction to the rate
of star formation of a factor $\sim 5$ at $z=3$ and $\sim 15$ at $z=4$. The
total stellar mass density today is $\rho_s(0)=5.0\times 10^8\mdden$
($\Omega_sh_{50}^2=$0.007). 

Overall, the fit to the data is still acceptable, showing how the blue and
near-IR light at $z<1$ are {\it  relatively poor indicators of the star
formation history at early epochs}. The reason for this is the short timescale
available at $z\gta 2$, which makes the present-day stellar mass density rather
insensitive to a significant boost of the stellar birthrate at high redshifts.
By contrast, variations in the global SFR around $z\sim1.5$, where the
bulk of the stellar population was assembled, have a much larger impact.
The adopted extinction-redshift relation, in fact, implies negligible 
reddening at $z\lta 1$. Relaxing this -- likely unphysical -- assumption 
would cause the model to significantly overproduce the $K$-band local 
luminosity density. We have also
checked that a larger amount of hidden star formation at early epochs, as
recently advocated by Meurer \etal (1997), would generate too much blue, 1
\micron\, and 2.2 \micron\ light to be still consistent with the observations.
An IMF which is less rich in massive stars would only exacerbate
the discrepancy. 

\begin{figure}
\plotone{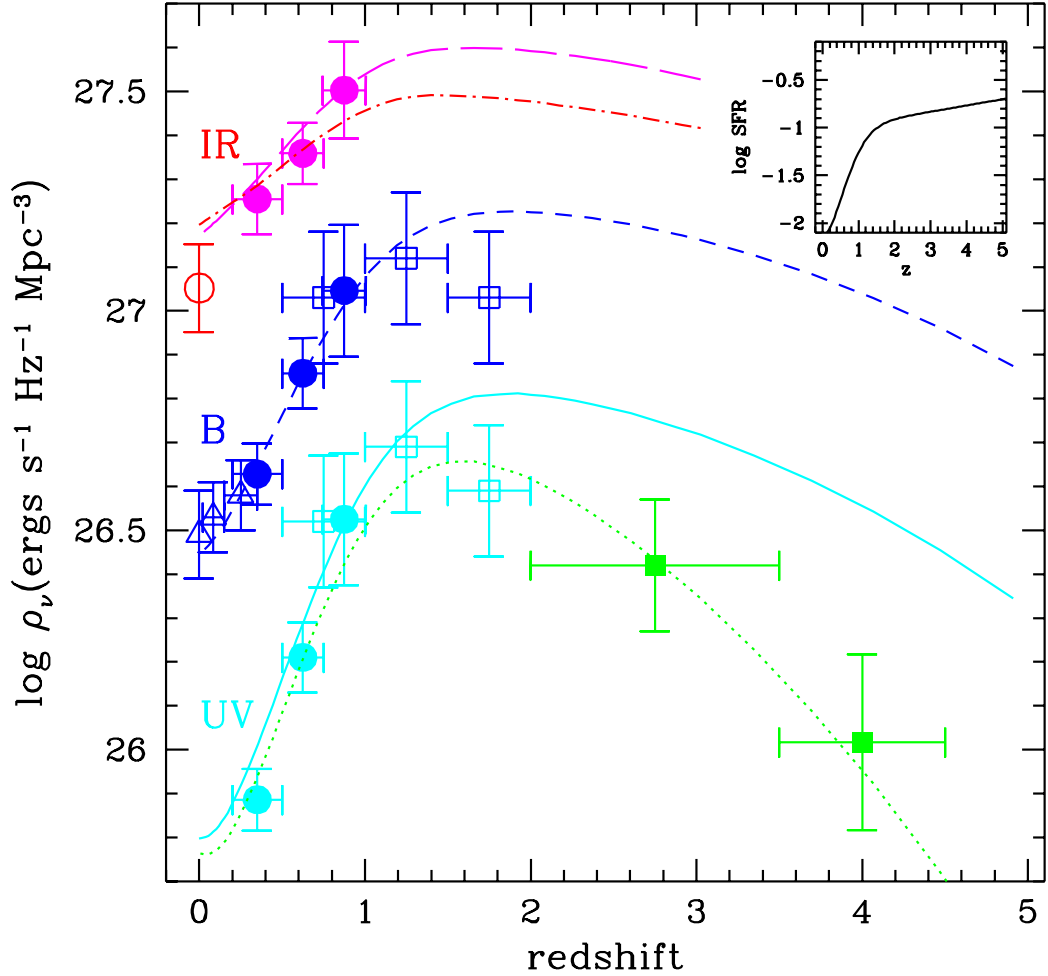}
\caption{Test case with a much larger star formation density at high redshift
than indicated by the HDF dropout analysis. The model -- designed to mimick a
``monolithic collapse'' scenario -- assumes a Salpeter IMF and a dust opacity
which increases rapidly with redshift, $E(B-V)=0.011(1+z)^{2.2}$. Notation is
the same as in Figure 3. 
\label{fig7}}
\end{figure}

\subsection{The Colors of High-Redshift Galaxies}

Figure 8 shows a comparison between the HDF data and the model predictions for
the evolution of galaxies in the $\ub$ vs. $\vi$ color-color plane according to
the star formation histories of Figures 3 and 4. The HDF ultraviolet passband
-- which is bluer than the standard ground-based $U$ filter -- permits the 
identification of star-forming galaxies in the interval $2\lta z\lta 3.5$. 
Galaxies in this redshift range predominantly occupy the top left portion of
the $\ub$ vs. $\vi$ color-color diagram because of the attenuation by the 
intergalactic medium and intrinsic absorption (M96). Galaxies at lower redshift
can have similar $\ub$ colors, but are typically either old or dusty, and are
therefore red in $\vi$ as well. The fact that the Salpeter 
IMF, $E(B-V)=0.1$ model reproduces quite well the rest-frame UV colors of
high-$z$ galaxies, while a dust-free $x=1.7$ IMF generates $\vi$ colors that
are 0.2 mag too blue, suggests the presence of some amount of dust extinction
in Lyman-break galaxies at $z\sim 3$ (Meurer \etal 1997; Dickinson \etal
1997). Adopting the greyer extinction law deduced by Calzetti \etal (1994)
from the integrated spectra of nearby starbursts would require larger 
corrections to the SFR at high-$z$ in order to match the observed 
colors. The consequence of this, however, would be the overproduction 
of red light at low redshifts, as noted in \S~5.4.  
Redder spectra can also result from an aging population or an IMF
which is less rich in massive stars than the adopted ones. 

\begin{figure}
\plotone{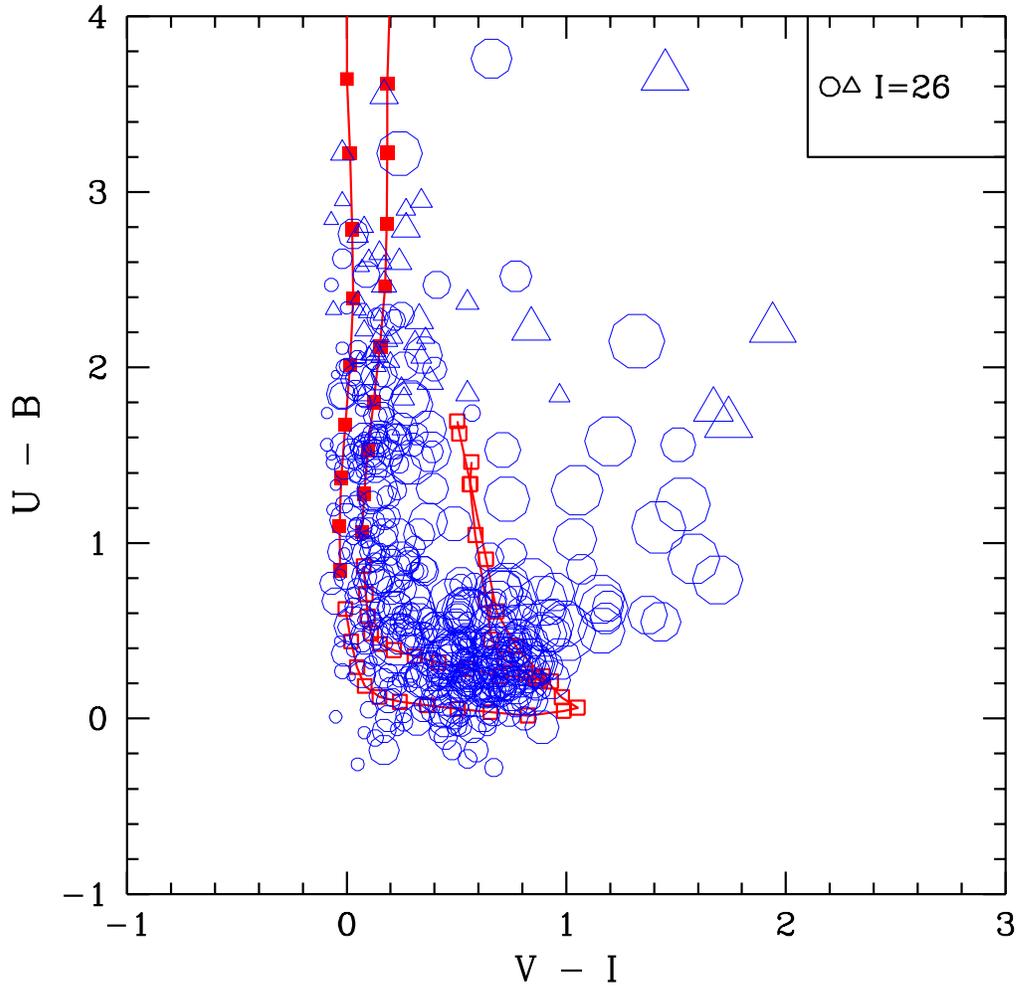}
\caption{{\it Solid lines:} model predictions for the color evolution of 
galaxies according to the star formation histories of Figure 3 ({\it right
curve}) and 4 ({\it left curve}). The points ({\it filled squares} for $z>2$
and {\it empty squares} for $z<2$) are plotted at redshift interval $\Delta
z=0.1$. {\it Empty circles} and {\it triangles:} colors of galaxies in the HDF
with $22<B<27$. Objects undetected in $U$ (with $S/N<1$) are plotted as
triangles at the 1$\sigma$ lower limits to their $U-B$ colors. Symbols size
scales with the $I$ mag of the object, and all magnitudes are given in the 
AB system. The ``plume'' of reddened high-$z$ galaxies is clearly seen in the
data. 
}
\label{fig8} 
\end{figure}

\subsection{Constraints from the Mid- and Far-Infrared Background}

Ultimately, it should be possible to set some constraints on the total amount
of star formation that is hidden by dust over the entire history of the
universe by looking at the cosmic infrared background (CIB) (Fall \etal 1996;
Burigana \etal 1997; Guiderdoni \etal 1997).  Studies of the CIB provide
information which is complementary to that given by optical observations. If
most of the star formation activity takes place within dusty gas clouds, the
starlight which is absorbed by various dust components will be reradiated 
thermally at longer wavelengths according to characteristic IR spectra. The
energy in the CIB would then exceed by far the entire background optical light
which is recorded in the galaxy counts. 

From an analysis of the smoothness
of the {\it COBE} DIRBE maps, Kashlinsky \etal (1996) have recently set an
upper limit to the CIB of 10--15 nW m$^{-2}$ sr$^{-1}$ at $\lambda=$10--100
\micron\ assuming clustered sources which evolve according to typical
scenarios. An analysis using data from {\it COBE} FIRAS by Puget \etal (1996)
(see also Fixsen \etal 1996) has revealed an isotropic residual at a level of 3.4 ($\lambda/400
\micron)^{-3}$ nW m$^{-2}$ sr$^{-1}$ in the 400--1000 \micron\ range, which
could be the long-searched CIB. The detection, recently revisited by Guiderdoni
\etal (1997), should be regarded as uncertain since it depends critically on
the subtraction of foreground emission by interstellar dust. 

By comparison, the total amount of starlight that is absorbed by dust and
reprocessed in the infrared is 7.5 nW m$^{-2}$ sr$^{-1}$ in the model depicted
in Figure 3, about 30\% of the total radiated flux. The monolithic collapse
scenario of Figure 7 generates 6.5 nW m$^{-2}$ sr$^{-1}$ instead. The resulting
CIB spectrum is expected to be rather flat because of the spread in the dust
temperatures -- cool dust will likely dominate the long wavelength emission,
warm small grains will radiate mostly at shorter wavelengths --  and the 
distribution in redshift. While both models appear then to be
consistent with the data (given the large uncertainties associated with the 
removal of foreground emission and with the observed and predicted spectral
shape of the CIB), it is clear that too much infrared light would be generated
by scenarios that have significantly larger amount of hidden star formation at 
early and late epochs.

\subsection{Metal Production}

We may at this stage use our set of models to establish a cosmic
timetable for the production of heavy elements (with atomic number $Z\ge 6$) in
relatively bright field galaxies (see M96). What we are interested in here is
the universal rate of ejection of newly synthesized material. In the
approximation of instantaneous recycling, the metal ejection rate per unit
comoving volume can be written as 
\begin{equation}
{\dot \rho_Z}=y(1-R)\times {\rm SFR}, \label{eq:rhoz}
\end{equation}
where the {\it net}, IMF-averaged yield of returned metals is 
\begin{equation}
y={\int mp_{\rm zm}\phi(m)dm\over (1-R)\int m\phi(m)dm},
\end{equation}
$p_{\rm zm}$ is the stellar yield, i.e., the mass fraction of a star of
mass $m$ that is converted to metals and ejected, and the dot denotes
differentiation with respect to cosmic time. 

The predicted end-products of stellar evolution, particularly from massive
stars, are subject to significant uncertainties. These are mainly due to the
effects of initial chemical composition, mass-loss history, the mechanisms of
supernova explosions, and the critical mass, $M_{\rm BH}$, above which stars
collapse to black holes without ejecting heavy elements into space. To be
quantitative, let us compare the nucleosynthetic enrichment from massive stars
($m\ge 9\msun$) according to the chemical yields tabulated by various authors.
For a Salpeter IMF, the total amount of freshly produced metals expelled in 
winds and final ejecta (in supernovae or planetary nebulae) gives, according
to Maeder (1992), $y=0.022$ for initial solar metallicity, high mass-loss
rates, and $M_{\rm BH}=120\msun$. When mass loss is small, as it is believed to
be the case for low metallicities, the net yield increases to $y=0.027$. The
Type II stellar yields without mass loss tabulated by Woosley \& Weaver (1995)
give $y=0.014$ (with $M_{\rm BH}=40\msun$ and $Z=Z_\odot$), those of Tsujimoto
\etal (1995) $y=0.024$ (with $M_{\rm BH}=70\msun$ and $Z=Z_\odot$). These
values are very sensitive to the choice of the IMF slope and lower-mass cutoff.
For a Scalo IMF in the assumed mass range ($0.1<M<125\msun$), the net
yield is typically a factor of 3.3 lower than Salpeter, and a 
factor of 4.5 lower for $x=1.7$. At the same time, a lower cutoff of
$0.5\msun$ would boost the net yield by a factor of 1.9 for Salpeter, 1.7 for
Scalo, and 3.1 for $x=1.7$. Note that some of these ambiguities partially 
cancel out when computing the total metal ejection rate, as the product
$y\times {\rm SFR}$ is less sensitive to the slope of the IMF than the yield or
the rate of star formation, and is insensitive to the lower mass cutoff.
Observationally, the best-fit ``effective yield'' (derived assuming a closed
box model) is 0.025$Z_\odot$ for Galactic halo clusters, 0.3$Z_\odot$ for disk
clusters, 0.4$Z_\odot$ for the solar neighborhood, and 1.8$Z_\odot$ for the
Galactic bulge (Pagel 1987). The last value may represent the universal true
yield, while the lower effective yields found in the other cases may be due,
e.g., to the loss of enriched material in galactic winds. 

Figure 9 shows the total mass of metals ever ejected, $\rho_Z$, versus 
redshift, i.e., the sum of the heavy elements stored in stars and in the gas
phase as given by the integral of equation (\ref{eq:rhoz}) over cosmic time.
The values plotted have been computed from the star formation histories
depicted in Figures 3 and 7, and have been normalized to $y\rho_s(0)$,
the mass density of metals at the present epoch according to
each model. A characteristic feature of the two competing scenarios is the
rather different average metallicity expected at high redshift. For comparison,
we have also plotted the {\it gas metallicity}, $Z_{\rm DLA}/Z_\odot$, as
deduced from observations by Pettini \etal (1997) of the damped Lyman-$\alpha$
systems (DLAs).  At early epochs,  when the gas consumption into stars is still
low, the metal mass density predicted from these models gives, in a closed box
model, a measurement of the metallicity of the gas phase. If DLAs and
star-forming field galaxies have the same level of heavy element enrichment,
then one would expect a rough agreement between $Z_{\rm DLA}$ and the model
predictions at $z\gta 3$. This is not true at $z\lta 2$, when a significant
fraction of heavy elements is locked into stars.\footnote{More complex chemical
evolution models which reproduce the metal enrichement history of the DLAs can
be found in Pei \& Fall (1995).}~ Without reading too much into
this comparison, it does appear that the monolithic collapse scenario tends to
overpredict the cosmic metallicity at high redshifts as sampled by the DLAs. 
In order for such a model to be acceptable, the gas traced by DLAs would
have to be physically distinct from the luminous star formation regions 
observed in the Lyman-break galaxies, and to be substantially under-enriched 
in metals compared to the cosmic mean.

It has been recently pointed out by Renzini (1997) and Mushotzky \& Loewenstein
(1997) that, in the absence of any systematic cluster/field differences,
clusters of galaxies may also provide an indication of the metal formation
history of the universe. In the Salpeter IMF model of Figure 3, the global
metallicity of the local universe is $y\Omega_s/\Omega_b\approx 0.1
y/Z_\odot$ solar, to be compared with the overall cluster metal abundance,
$\sim 1/3$ solar. If $y\sim Z_\odot$, the efficiency of metal production must
have been larger in clusters than in the field, in spite of both having a
similar baryon-to-star conversion efficiency, $\Omega_s/\Omega_b\sim 10\%$
(Renzini 1997). Alternatively, a yield $y\sim 3 Z_\odot$ may solve the apparent
discrepancy. In this case, field galaxies would have to have ejected a
significant amount of the heavy elements they produced, and there should be a
comparable share of metals in the intergalactic medium (IGM) as there is in the
intracluster gas. 

\begin{figure}
\plotone{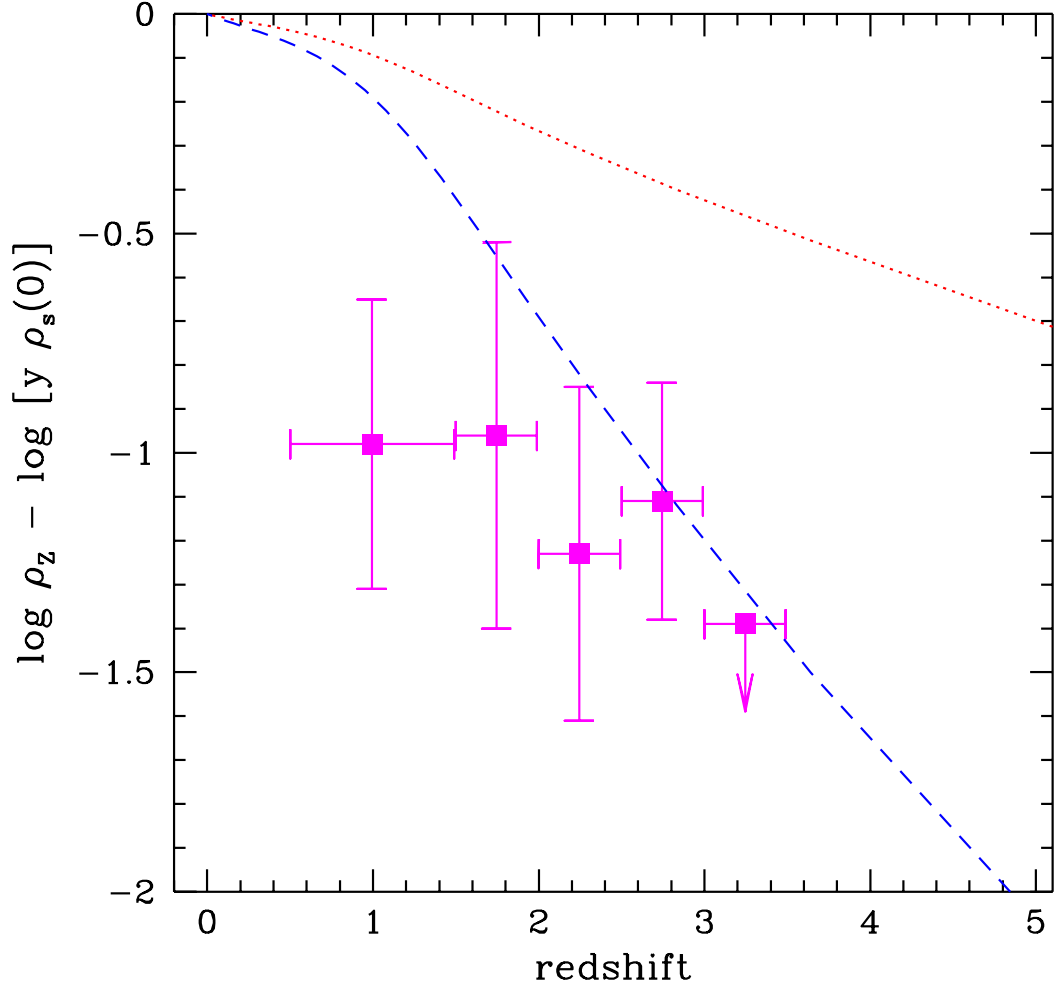}
\caption{Total mass of heavy elements ever ejected versus redshift for the
Salpeter IMF model of Figure 3 ({\it dashed line}) and the ``monolithic
collapse'' model of Figure 7 ({\it dotted line}), normalized to $y\rho_s(0)$,
the total  mass density of metals at the present epoch. {\it Filled squares:}
column density-weighted metallicities (in units of solar) as derived from
observations of the damped Lyman-$\alpha$ systems (Pettini \etal 1997). 
\label{fig9}}
\end{figure}

\acknowledgments
We have benefited from many stimulating discussions with G. Bruzual, D.
Calzetti, S. Charlot, M. Fall, N. Panagia, M. Pettini, and A. Renzini on
various topics related to this work. We are indebted to G. Bruzual and S.
Charlot for providing the newer population synthesis models, to A. Connolly for
computing the $B$-band luminosity density from his photometric redshift
catalog, and to M. Treyer and R. Ellis for communicating their unpublished
results on the local UV luminosity function. We thank the referee, Raja
Guhathakurta, for carefully reading the manuscript and for valuable comments.
Support for this work was provided by NASA through grants NAG5-4236 and
AR-06337.10-94A from the Space Telescope Science Institute, which is operated
by the Association of Universities for Research in Astronomy, Inc., under NASA
contract NAS5-26555. 

\clearpage 
\begin{deluxetable}{lccccccccr}
\scriptsize
\tabcolsep 0.0pt
\tablewidth{6.0in}
\tablenum{1}
\tablecaption{Comoving Luminosity Density\tablenotemark{a} \label{lumden}}
\tablehead{
\colhead{redshift} & \colhead{1500 \AA} & \colhead{2800 \AA}
& \colhead{4400 \AA} & \colhead{1 \micron}& \colhead{2.2 \micron} 
}
\startdata
\cutinhead{Gardner \etal (1997)}
0.0  & \nodata & \nodata  & \nodata  & \nodata  &  27.05$\pm$0.1\nl
\nl
\cutinhead{Lilly \etal (1996)}
0.20--0.50 & \nodata  & 25.89$\pm$0.07 & 26.63$\pm$0.07 & 27.26$\pm$0.08  & \nodata \nl
0.50--0.75 & \nodata  & 26.21$\pm$0.08 & 26.86$\pm$0.08 & 27.36$\pm$0.07  & \nodata \nl 
0.75--1.00 & \nodata  & 26.53$\pm$0.15 & 27.05$\pm$0.13 & 27.50$\pm$0.11  & \nodata \nl
\nl
\cutinhead{Ellis \etal (1996)}
0.0        & \nodata  & \nodata  &  26.49$\pm$0.10 & \nodata  & \nodata \nl
0.02--0.15 & \nodata  & \nodata  &  26.53$\pm$0.08 & \nodata  & \nodata \nl 
0.15--0.35 & \nodata  & \nodata  &  26.58$\pm$0.08 & \nodata  & \nodata \nl
\nl
\cutinhead{Connolly \etal (1997)}
0.50--1.00 & \nodata  & 26.52$\pm$0.15  & 27.03$\pm$0.15& \nodata  & \nodata  \nl 
1.00--1.50 & \nodata  & 26.69$\pm$0.15  & 27.12$\pm$0.15& \nodata  & \nodata  \nl
1.50--2.00 & \nodata  & 26.59$\pm$0.15  & 27.03$\pm$0.15& \nodata  & \nodata \nl
\nl
\cutinhead{Madau \etal (1996) and Madau (1997a)}
2.00--3.50 & 26.42$\pm$0.15 & \nodata  & \nodata  & \nodata  & \nodata \nl 
3.50--4.50 & 26.02$\pm$0.20 & \nodata  & \nodata  & \nodata  & \nodata \nl 
\nl
\enddata
\tablenotetext{a}{Values listed are log luminosity density in units of
$\ldunits$.}
\end{deluxetable}
\clearpage

\references

Babul, A., \& Ferguson, H.~C. 1996, \apj, 458, 100
 
Baugh, C.~M., Cole, S., Frenk, C.~S., \& Lacey, C.~G. 1997, \apj, submitted

Bower, R.~G., Lucey, J.~R., \& Ellis, R.~S. 1992, \mnras, 254, 589

Broadhurst, T.~J., Ellis, R.~S., \& Glazebrook, K. 1992, Nature, 355, 55

Bruzual, A.~G., \& Charlot, S. 1993, \apj, 405, 538

Bruzual, A.~G., \& Charlot, S. 1997, in preparation

Bruzual A.~G., \& Kron, R.~G. 1980, \apj, 241, 25 

Burigana, C., Danese, L., De Zotti, G., Franceschini, A., Mazzei, P., 
\& Toffolatti, L. 1997, \mnras, 287, L17

Caldwell, J.~A.~R., \& Ostriker, J.~P. 1981, \apj, 251, 61

Calzetti, D., Kinney, A.~L., \& Storchi-Bergmann, T. 1994, \apj, 429, 582

Carlberg, R.~G., \& Charlot, S. 1992, \apj, 397, 5

Charlot, S. 1996a, in From Stars to Galaxies, eds. C. Leitherer \& U. 
Fritze-von Alvensleben (ASP Conference Series), in press

Charlot, S. 1996b, in The Universe at High-z, Large Scale Structure, and 
the Cosmic Microwave Background, eds. E. Martinez-Gonzalez \& J.~L. Sanz
(Heidelberg: Springer), p. 53

Charlot, S., Worthey, G., \& Bressan, A. 1996, \apj, 457, 625

Cole, S., Arag\'on-Salamanca, A., Frenk, C.~S., Navarro, J.~F., \&
Zepf, S.~E. 1994, \mnras, 271, 781

Connolly, A.~J., Szalay, A.~S., Dickinson, M.~E., SubbaRao, M.~U., 
\& Brunner, R.~J. 1997, \apj, in press

Cowie, L.~L., Songaila, A., Hu, E.~M., \& Cohen, J.~G. 1996, \aj, 112, 839

Dickinson, M.~E. 1997, private communication

Dickinson, M.~E., \etal 1997, in preparation

Eggen, O.~J., Lynden-Bell, D., \& Sandage, A.~R. 1962, \apj, 136, 748

Ellis, R.~S. 1997, ARA\&A, 35, in press

Ellis, R.~S., Colless, M., Broadhurst, T., Heyl, J., \& Glazebrook,
K. 1996, \mnras, 280, 235

Fall, S.~M., Charlot, S., \& Pei, Y. C. 1996, \apjl, 464, L43

Ferguson, H.~C. 1997, in The Hubble Space Telescope and the High Redshift 
Universe, eds. N. R. Tanvir, A. Arag{\'o}n-Salamanca, \& J. V. Wall, 
(Singapore: World Scientific Press), in press

Ferguson, H.~C., \& Babul, A. 1997, \mnras, in press

Fixsen, D.~J., Cheng, E.~S., Gales, J.~M., Mather, J.~C., Shafer, R.~A., 
Wright, E.~L. 1996, \apj, 473, 576

Gallego, J., Zamorano, J., Arag{\'o}n-Salamanca, A., \& Rego, M. 1995, 
\apjl, 455, L1

Gardner, J.~P., Sharples, R.~M., Frenk, C.~S., \& Carrasco, B.~E. 1997,
\apjl, 480, L99

Gronwall, C., \& Koo, D.~C. 1995, \apjl, 440, L1

Guiderdoni, B., Bouchet, F.~R., Puget, J.-L., Lagache, G., \& Hivon, E.
1997, Nature, in press

Guiderdoni, B., \& Rocca-Volmerange, B. 1990, A\&A, 227, 362

Kashlinsky, A., Mather, J.~C., \& Odenwald, S. 1996, \apjl, 473, L9

Kauffmann, G., \& White, S.~D.~M. 1993, \mnras, 261, 921

Kauffmann, G., White, S.~D.~M., \& Guiderdoni, B. 1993, \mnras, 264, 201

Koo, D.~C. 1985, \aj, 90, 418

Lacey, C.~G., \& Silk, J. 1991, \apj, 381, 14

Lilly, S.~J., Le F{\'e}vre, O., Hammer, F., \& Crampton, D., 1996, \apjl, 460,
L1 

Lilly, S.~J., Tresse, L., Hammer, F., Crampton, D., \& Le F{\'e}vre, O. 1995,
\apj, 455, 108 

Madau, P. 1997a, in Star Formation Near and Far, eds. S. S. Holt \& G. L.
Mundy, (AIP: New York), p. 481

Madau, P., Ferguson, H.~C., Dickinson, M.~E., Giavalisco, M., 
Steidel, C.~C., \& Fruchter, A. 1996, \mnras, 283, 1388 (M96)

Maeder, A. 1992, A\&A, 264, 105

McGaugh, S.~S., \& Bothun, G.~D. 1994, \aj, 107, 530

Metcalfe, N., Shanks, T., Fong, R., \& Jones, L.~R. 1991, \mnras, 249, 498

Meurer, G.~R., Heckman, T.~M., Lehnert, M.~D., Leitherer, C., \& Lowenthal, J.
1997, \aj, 114, 54

Milliard, B., Donas, J., Laget, M., Armand, C., \& Vuillemin, A. 1992, A\&A,
257, 24 

Mushotzky, R.~F., \& Loewenstein, M. 1997, \apj, 481, L63

Ortolani, S., Renzini, A., Gilmozzi, R., Marconi, G., Barbuy, B., Bica, E., \&
Rich, M.~R. 1995, Nature, 377, 701 

Pagel, B.~E.~J. 1987, in The Galaxy, eds. G. Gilmore \& B. Carswell
(Reidel: Dordrecht), p. 341

Pei, Y.~C. 1992, \apj, 395, 130

Pei, Y.~C., \& Fall, S.~M. 1995, \apj, 454, 69

Pettini, M., Smith, L.~J., King, D.~L., \& Hunstead, R.~W. 1997, \apj, in press

Pozzetti, L., Bruzual, G.~A., \& Zamorani, G. 1996, \mnras, 281, 953

Pozzetti, L., Madau, P., Ferguson, H.~C., Zamorani, G., \& Bruzual, G.~A. 1997,
\mnras, submitted 

Puget, J.-L., Abergel, A., Bernard, J.-P., Boulanger, F., Burton, W.~B., Desert,
F.-X., \& Hartmann, D. 1996, A\&A, 308, L5

Renzini, A. 1995, in Stellar Populations, ed. P.C. van der Kruit \& G. Gilmore
(Dordrecht: Kluwer), p. 325

Renzini, A. 1997, \apj, in press

Salpeter, E.~E. 1955, \apj, 121, 161

Saunders, W., Rowan-Robinson, M., Lawrence, A., Efstathiou, G., Kaiser, N.,
Ellis, R.~S., \& Frenk, C. S. 1990, \mnras, 242, 318 

Scalo, J.~N. 1986, Fundam. Cosmic Phys., 11, 1

Schechter, P.~L., \& Dressler, A. 1987, \aj, 94, 56

Searle, L., \& Zinn, R. 1978, \apj, 225, 357

Soifer, B.~T., \& Neugebauer, G. 1991, \aj, 101, 354

Tinsley, B.~M. 1980, Fundam. Cosmic Phys., 5, 287

Tinsley, B.~M., \& Gunn, J.~E. 1976, \apj, 203, 52

Treyer, M.~A., Ellis, R.~S., Milliard, B., \& Donas, J. 1997, in 
The Ultraviolet Universe at Low and High Redshift, ed. W. Waller,
(Woodbury: AIP Press), in press

Tsujimoto, T., Nomoto, K., Yoshii, Y., Hashimoto, M., Yanagida, S., \&
Thielemann, F.-K. 1995, \mnras, 277, 945 

Weidemann, V. 1987, A\&A, 188, 74

White, S.~D.~M., \& Frenk, C.~S. 1991, \apj, 379, 25

Williams, R.~E., \etal 1996, \aj, 112, 1335

Woosley, S.~E., \& Weaver, T.~A. 1995, \apjs, 101, 181

\end{document}